\documentclass[12pt]{JHEP3}
\usepackage{psfig}
\usepackage{epsfig}
\usepackage{citesort}       
\usepackage{multicol}
\def\Red{}
\def\Black{}

\def\lsim{\raise0.3ex\hbox{$\;<$\kern-0.75em\raise-1.1ex\hbox{$\sim\;$}}}
\def\gsim{\raise0.3ex\hbox{$\;>$\kern-0.75em\raise-1.1ex\hbox{$\sim\;$}}}
\newcommand{\vecc}[1]{\mbox{\boldmath $#1$}}
\newcommand{\vR}{\vecc{R}}            

\newcommand{\matr}[1]{\mbox{#1}}
\newcommand{\eV}{\matr{eV} }

\def\CCSNO{\mbox{\rm SNO-CC} }
\def\ESSNO{\mbox{\rm SNO-ES} }
\def\NCSNO{\mbox{\rm SNO-NC} }
\newcommand{\boron}{\mbox{${}^8$B\ }}
\newcommand{\fluxunit}{\mbox{$10^6$ cm$^{-2}$ s$^{-1}$ }}

\def\npbps#1#2#3{  { Nucl. Phys. }(Proc. Suppl.){\bf B #1} (19#2) #3}
\def\plb#1#2#3{    { Phys. Lett. }{\bf B #1} (19#2) #3}

\def\prl#1#2#3{    { Phys. Rev. Lett. }{\bf #1} (19#2) #3}

\preprint{\small
CERN-TH-2003-150; IFUM-746/FT\\
FT-UM-TH-03-12; IFTM-03-10\\
ULB-TH/03-31
}
\title{\Red The Neutrino mass matrix after 
 Kamland and SNO salt enhanced results\Black}
\author{ P. Aliani$^{a}$, V. Antonelli$^{b,c}$, 
M. Picariello$^{b,c}$, E. Torrente-Lujan$^{d,e,f}$\\
{\small\sl
$^{a}$ Service de Physique Th\'eorique, Universit\'e Libre  Bruxelles, Bruxelles, Belgium\\
$^{b}$ Dip. di Fisica, Universit\`a degli Studi di Milano, Milano, Italy\\ 
$^{c}$ I.N.F.N., Sezione di Milano, Milano, Italy\\
$^{d}$ Dept. de Fisica, Universidad de Murcia, Murcia, Spain\\
$^{e}$ Instituto de Fisica Teorica, Universidad de Murcia, Murcia, Spain\\
$^{f}$ CERN-TH Division, CH-1211 Geneve 23, Switzerland\\
\email{paola.aliani@ulb.ac.be, vito.antonelli@mi.infn.it,
 marco.picariello@mi.infn.it, torrente@cern.ch}}
}
\abstract{
An updated analysis of all available neutrino oscillation evidence 
 in Solar experiments  including the latest $SNO ES, CC$ and $NC$ data 
(254d live time, NaCL enhanced efficiency) is presented. We obtain, for the 
fraction of active oscillating neutrinos:
$\sin^2\alpha=(\Phi_{NC}-\Phi_{CC})/(\Phi_{SSM}-\Phi_{CC})=0.940^{+0.065}_{-0.060}$
nearly $20\sigma$  from the pure sterile oscillation case. The fraction of 
oscillating sterile neutrinos  $\cos^2\alpha \lsim 0.12$ (1$\sigma$ CL).
At face value, these results  might slightly favour the existence of a
small sterile oscillating sector. In the framework of two active
neutrino oscillations we determine 
individual neutrino mixing parameters and their errors and we obtain 
$ \Delta m^2= 7.01\pm 0.08\times 10^{-5} \eV^2,\quad 
\tan^2\theta=  0.42^{+0.12}_{-0.07}.$ 
The main difference with previous analyses is a better resolution in
parameter space. 
In particular the secondary region at larger mass differences (LMAII)
is now excluded 
at 95\% CL. The combined analysis of solar and Kamland data concludes that 
maximal mixing is not favoured at $\sim 4-5\sigma$. This  is not
supported by the 
antineutrino reactor  results alone. We also estimate the individual
elements of the two neutrino mass matrix, writing  $M^2=m^2 I+M_0^2$,
we obtain ($1\sigma$ errors):
$$M_0^2=10^{-5}\ eV^2 \pmatrix{ 2.06^{+0.29}_{-0.31} & 3.15^{+0.29}_{-0.35}
\cr  3.15^{+0.29}_{-0.35} & 4.60^{+0.56}_{-0.44}}.$$
}
\begin{document}
\section{Introduction}
The Sudbury Neutrino Observatory (SNO) collaboration has recently presented 
data relative to the NaCl phase of the experiment \cite{newSNO}. 
The addition of NaCl to a pure D$_{2}$O detection medium 
increases the detector's sensitivity to the neutral-current (NC) reactions 
within its fiducial volume, enhancing the NC detection 
efficiency of about a factor three with respect to the previous `no-salt' 
phase.
This new result adds to a list of successive, compatible, ever more accurate 
results. 
The year 2002 was fundamental for neutrino physics and many long standing 
puzzles found their final solution. 
The data published in springtime by SuperKamiokande 
\cite{Fukuda:2002pe,Smy:2002fs} and SNO \cite{Ahmad:2002ka,Ahmad:2002jz} 
and the first results from the reactor experiment KamLAND \cite{Eguchi:2002dm}
in December, confirmed the evidence accumulated in about 30 years of
solar neutrino experiments, proved in a crystal clear way that
neutrinos are massive and oscillating particles and selected the
so-called Large Mixing Angle (LMA) between the different possible
solutions of the Solar Neutrino Problem (SNP).

After the publication of the first KamLAND data a new era has started,
in which the main issue is no longer to prove the validity of the
oscillation hypothesis, but rather to determine with the best possible
accuracy the oscillation parameters entering the PMNS mixing matrix.
As it has been pointed out in previous papers (see for
instance~\cite{Migliozzi:2003pw,Bandyopadhyay:2003du,Bahcall:2003ce,Choubey:2003qx}),
the accurate knowledge of the  "solar mixing parameters''
($\theta_{12}$ and $\Delta m_{12}^2$ in the case of normal hierarchy)
is essential not only for a better comprehension of this sector of
neutrino physics, but also for the future study of the other mixing
parameters.

In fact the neutrino parameters drive the subdominant effects in the
long baseline and neutrino factory experiments and
a lack of information about their exact values could seriously reduce
the accuracy in the determination of the other elements of the mixing
matrix and  eventually of the CP violation parameter $\delta$.
Hence, improving the present accuracy in the determination of the solar mixing
parameters is mandatory not only to understand the physical potential of 
future experiments but also in order to optimize their set-up in the best 
possible way. 

From this point of view, the situation was still quite unsatisfactory
after the publication of the first KamLAND data.
The many different analysis of these
data~\cite{klothers,Aliani:2002na,Antonelli:2002zf} 
agree on the fact that these results, together with all the solar neutrino
experiments, essentially select two distinct
regions in the solar mixing parameter plan, both corresponding to the
LMA solution. These two regions are usually denoted as ``low LMA" and
``high LMA". The mixing angle was not maximal, even if it was large
and it was constrained 
in the interval $0.27 \leq \tan^2 \theta_{12} \leq 0.88$ at the
3$\sigma$ level.
One can see  that the uncertainty on the value of the mixing angle was still 
quite significant. Moreover it was not possible to discriminate between the 
two different possibilities for the mass difference.
It is clear that a step further in the direction of pinning down more accurately the mixing parameters was needed.

The $SNO$ experiment measures the \boron Solar neutrinos via the
reactions~\cite{SNO,Boger:2000bb,Barger:2001pf,Bahcall:2001hv,Bahcall:2001hv}:
1) Charged Current (CC): $\nu_e + d\rightarrow 2 p+e^-$,
2) Elastic Scattering (ES): $\nu_x + e^-\rightarrow \nu_x+e^-$.
3) Neutral Current(NC): $\nu_x + d\rightarrow p+n+\nu_x$.
The first reaction is sensitive exclusively to electron neutrinos.
The second, the same used in SuperKamiokande (SK),
 is instead sensitive, with different efficiencies, to all flavors.
Finally the NC reaction is equally sensitive to all active neutrino 
species.
Hence SNO can measure simultaneously the electron and active non-electron 
neutrino component of the solar flux at high energies ($\gsim $5 MeV).
The novelty of the `salt phase' is that it is now possible to distinguish 
clearly the NC events from the ES and CC ones and therefore to 
analyse the data without making use of the no-spectrum-distortion hypothesis
\cite{newSNO,newsnoothers}.

The non-electron component is found to be $\sim 5\sigma $ greater than 
zero, the standard prediction, thus providing the strongest evidence so 
 far for flavour oscillation in the neutral lepton sector:
the agreement  of the total flux, provided by the NC measurement
 with the expectations 
 implies as a by-product the confirmation of the validity of the
 SSM~\cite{turck,bpb2001,bp95}.

In this work we present an up-to-date analysis of all available Solar 
neutrino evidence, including the latest global SNO
results~\cite{newSNO} (together with the data of previous SNO
phases~\cite{Ahmad:2002jz,Ahmad:2002ka,sno2001} 
and of the other
experiments~\cite{homestake,newgallium,Fukuda:2002pe}, and of the
KamLAND data~\cite{Eguchi:2002dm}.
A similar analysis using in addition the NaCl enhanced SNO spectrum will 
be presented elsewhere \cite{ussnospec}.
 As a result, the new SNO data make the discrimination between the two
different $\Delta m^2$ regions possible. In particular the secondary
region at larger mass differences
secondary region at larger mass differences
(LMAII or high LMA) is now excluded at 95\% CL.
 The combined analysis of solar and KamLAND data continues to conclude
 that maximal mixing is not favored at an even larger significance
level as before.
Finally, we show that the achieved resolution in neutrino parameter
space is good enough for an 
estimation of the individual elements of two neutrino mass matrix and
their errors. 
 As it will be shown below we obtain a square mass matrix:
$$M_0^2= 10^{-5}\ eV^2 
\pmatrix{ 2.06^{+0.29}_{-0.31} & 3.15^{+0.29}_{-0.35}\cr  3.15^{+0.29}_{-0.35} & 4.60^{+0.56}_{-0.44}}.$$
This result can be sharpened and a mass matrix can already be given if
a concrete value for the absolute neutrino mass scale is assumed. 

The structure of the present work is the following.
First (section 2) we update some model independent results which put in a 
quantitative basis  the extent of the deviations with respect to the 
 standard non-oscillating case and the relative importance of active/sterile 
 oscillations. After Section 3 dedicated to general description of methods,
 we determine (Section 4) the allowed areas in parameter space in the
framework of active two neutrino oscillations from a standard
statistical analysis. 
Individual  values for $\Delta m^2$ and $\tan^2\theta$ with error estimation 
are  obtained from the analysis of marginal likelihoods.

\section{Some Model Independent Results}

Different quantities can be defined in order to make  the evidence for 
disappearance and  appearance of the neutrino flavours explicit. 
From the three fluxes measured by SNO is possible to 
define two useful ratios, deviations of these ratios with
respect to  their 
standard value are powerful tests for occurrence of new physics.
Here we update the computations of Ref.\cite{Aliani:2002ma} for  the
values for $\Phi_{CC}/\Phi_{ES}$ and 
$\Phi_{CC}/\Phi_{NC}$  being extremely careful with the treatment of the 
correlations on the incertitudes.  The inclusion or not of these correlations
can affect significantly the results for these 
ratios (see table II in Ref.\cite{newSNO} for a complete
 list of systematical errors).
The results we obtain from the new SNO data are similar to the old
ones except for a strong decrease in the error bars in some cases.
From the value from SNO rates\cite{newSNO} we obtain
$$ \frac{\Phi_{CC}}{\Phi_{ES}}=0.691^{+0.150}_{-0.096}, $$
 a value which is  $\sim$ 2.1 $\sigma$ away from the  no-oscillation
expectation value of one. 
The ratio of CC and NC fluxes gives the fraction of electron neutrinos
remaining in the solar neutrino beam at detection point. We obtain
$$\frac{\Phi_{CC}}{\Phi_{NC}}=0.305^{+0.030}_{-0.024},$$
this value is nominally many standard deviations ($\sim 20 \sigma$)
away  from the  standard model case \cite{Bahcall:1996bw}.  

Finally, if in addition to SNO data  we consider the 
 flux predicted by the solar standard model one can obtain  
$\sin^2\alpha$, the fraction of 
 active oscillating neutrinos, again using 
 the SNO data and fully applying systematic correlations, we find:
\begin{eqnarray}
\sin^2\alpha&=&
\frac{\Phi_{NC}-\Phi_{CC}}{\Phi_{SSM}-\Phi_{CC}}=0.940^{+0.065}_{-0.060}\\
\cos^2\alpha &\lsim&  0.12\quad ( 1\sigma),
\end{eqnarray}
where the fraction of oscillating sterile neutrinos 
 $\cos^2\alpha\equiv 1-\sin^2\alpha$. 
The SSM flux is taken as the ${}^8\rm B$ flux predicted in the revised 
Ref.\cite{bpb2001}.
Although slightly increasing, the central value is still clearly below one 
(only-active oscillations case). 
Although electron neutrinos are still allowed to oscillate into 
sterile neutrinos the hypothesis of transitions to {\em only} sterile 
 neutrinos is rejected at nearly $15\sigma$.
 On the other hand, as a consequence of the reduced error bars, this data 
can be taken as a mildly positive hint in favour of a small sterile component:
the pure active case is now one sigma away from the central value (pure 
active oscillations are ``excluded'' at one sigma).

\section{Methods and statistical procedures}

The  computation  of the neutrino oscillation probabilities
 in Solar and Earth matter and of the expected signal in each experiment  
follows the standard methods found in the 
literature~\cite{Barger:2001zs,Fogli:2001vr,Krastev:2001tv,Bahcall:2001zu,Bandyopadhyay:2001fb,Choubey:2001bi,Bandyopadhyay:2001aa,torrente2001,torrente}. 
We  solve numerically \cite{torrente2001,Aliani:2003me},
the neutrino evolution equations  for all the oscillation parameter space.
The survival probabilities for an electron neutrino, produced in the 
 Sun, to arrive at the Earth  are calculated in three steps.
The propagation from the production point to sun's surface is computed 
 numerically in all the parameter range 
 using the electron number density $n_e$ given by the 
 BPB2001 model~\cite{bpb2001} averaging over the production point.
The propagation in vacuum from the Sun surface to the Earth is computed 
 analytically. The averaging over the annual  variation of the orbit
 is also exactly performed using simple Bessel functions.
To take the Earth matter effects into account, we adopt a
 spherical model of the Earth  density and chemical composition.
In this model, the Earth is divided in eleven  radial density
 zones~\cite{earthprofile}, in each of which  a polynomial
 interpolation is used to obtain the electron density.
The composition of the neutrino propagation in the three different 
 regions is performed exactly using an evolution operator
 formalism~\cite{torrente}.
The final survival probabilities are obtained from the corresponding
 (non-pure) density matrices built from the evolution operators in 
 each of these three regions.
The  night quantities are obtained using appropriate weights 
 which depend on the neutrino  impact parameter and the 
 sagitta distance from neutrino trajectory  to the Earth center,
 for each detector's geographical location.
In this analysis in addition to night probabilities we will 
need the  partial night probabilities corresponding 
to the 6 zenith angle bin data presented by SK \cite{Smy:2002fs}.
They are obtained using appropriate weights 
 which depend on the neutrino  impact parameter and the 
 sagitta distance from neutrino trajectory  to the Earth's center,
 for each detector's geographical location.

The expected signal in each detector is obtained by convoluting neutrino 
 fluxes, oscillation probabilities, neutrino cross sections and detector 
 energy response functions. We have used 
 neutrino-electron elastic cross sections which include radiative
 corrections~\cite{sirlin1994}.
Neutrino cross sections on deuterium needed for the computation of the 
 SNO measurements are taken from~\cite{nakamura2001}.
Detector effects are summarized by the respective response functions,
 obtained by taking into account both the energy resolution and the detector
 efficiency.
We obtained the energy resolution function for SK using the data
 presented in~\cite{Nakahata:1999pz,skthesis,sakurai}. 
The effective threshold efficiencies, which take into account the live 
 time for each experimental period, are incorporated into our
 simulation program. They are obtained from~\cite{Fukuda:2001nj}. 
The resolution function and other characteristics for SNO used here are 
 those given in~Refs.\cite{newSNO,Ahmad:2002ka,Ahmad:2002jz,snohowto}.

%
%

The two principal ingredients in the calculation of the expected 
signal in KamLAND are  the reactor  flux and
the antineutrino cross section on protons. 
We refer to Ref.\cite{Aliani:2002na} for a detailed account of the 
methods used here.
In summary, in order to obtain the expected number of events at KamLAND, 
we sum the expectations for all the relevant reactor sources weighting 
each source by its power and distance to the detector
(table II in Ref.~\cite{Murayama:2000iq} ), 
 assuming the same 
spectrum originated from each reactor. 
We sum over the nearby  power reactors, we neglect 
farther Japanese and Korean reactors and even farther 
rest-of-the-world reactors which  give only a minor additional
contribution.
The average expected signal in each energy bin
is given by the convolution of  the oscillation probability 
averaged  over the  distance and power of the different 
reactors.
Expressions for the 
antineutrino capture cross section  are taken from 
the literature \cite{vogel,kltorrente}. The  matrix 
element  for this cross section can 
be written in terms of the neutron half-life, 
 we have used the latest published 
value $t_{1/2}=613.9\pm 0.55$ \cite{PDG2002}.
The antineutrino flux spectrum, 
the relative reactor-reactor 
 power normalization which 
is included in the definition of 
 the effective probability and 
the energy resolution of KamLAND are used in addition. 
The  energy resolution in the prompt 
positron detection is 
 obtained by us from the raw calibration 
data presented in Ref.\cite{Aliani:2002na,klstony}.
Moreover, we assume a  408 ton fiducial 
mass and  standard 
nuclear plant power and fuel schedule, we take 
an averaged, time-independent, fuel composition 
equal for each detector. Detection efficiency is 
taken close  100\% and independent of the energy \cite{Eguchi:2002dm}.

\subsection{Statistical Analysis}

The statistical significance of the neutrino oscillation hypothesis
 is tested with a standard $\chi^2$ method which is explained in 
 detail in Ref.\cite{torrente2001}.
Our present analysis is based on the consideration of a
 $\chi^2$ 
sum of  two distinct contributions,  one coming from SNO global rates 
and all the rest of  solar neutrino data and the contribution of the 
KamLAND experiment $  \chi^2= \chi^2_{\odot} + \chi^2_{KL}$.
The contribution from KamLAND includes the  binned  signal
(See table 2 in  Ref.\cite{Aliani:2002na})  as is explained in 
detail in Ref.\cite{Eguchi:2002dm}. In summary, the KamLAND contribution is made of 
two parts, one with
$ \chi^2_{gl,KL}=  (R^{exp}-R^{th})^2/\sigma^2$.
The experimental signal $R^{exp}$ and statistical and systematic errors are 
 shown in Table~\ref{t1}.
 The contribution of the KamLAND spectrum is as follows:
\begin{eqnarray}
  \chi^2_{\rm spec,KL}&=& ({\alpha \vR^{\rm th}-\vR^{\rm exp}})^t 
\left (\sigma_{unc}^2+\sigma_{corr}^2 \right)^{-1} ({\alpha \vR^{\rm th}-\vR^{\rm exp}})
\label{chiklspec}
\end{eqnarray}
The total error matrix $\sigma$ is computed as a sum of assumed 
systematic deviations, $\sigma_{sys}/S\sim 6.5\%$,
 mainly coming from flux uncertainty 
($3\%$), energy calibration and threshold
(see table II of
 Ref.~\cite{Eguchi:2002dm}. for a total systematic error
$\sim 6.4\%$), see also Ref.\cite{klstony,Murayama:2000iq,kamlandmonaco}) 
and statistical errors. The parameter $\alpha$ is a free normalization parameter.
The   effect of systematic sources on 
individual bin deviations  has been computed  
by us studying the influence on the response function, 
furtherly 
we have assumed full correlation among bins.

The solar neutrino contribution can be written in 
the following way:
\begin{eqnarray}
  \chi^2_{\odot }&=& \chi^2_{\rm glob}+\chi^2_{\rm SK}+\chi^2_{\rm SNO}.
\end{eqnarray}
The function $\chi^2_{\rm glob}$ 
correspond to the total event rates measured at the  
Homestake  experiment~\cite{homestake} and at the gallium 
experiments SAGE~\cite{sage}, 
GNO~\cite{gno2000} and GALLEX~\cite{gallex}. 
We follow closely the definition used in 
previous works (see Ref.\cite{Aliani:2002ma}
 for definitions and 
 Table~(1) in Ref.\cite{Aliani:2002ma} for an explicit list of results 
and other  references).  
The contribution to the $\chi^2$ from the SuperKamiokande 
data ($\chi^2_{\rm SK}$) has been obtained by using  
  double-binned data in energy 
 and zenith angle (see table 2 in Ref.\cite{Smy:2002fs} and
 also Ref.\cite{Fukuda:2002pe}):
8 energy bins of variable width and 7 zenith angle bins which
 include the day bin and 6 night ones (see Ref.\cite{Eguchi:2002dm}).

The contribution of SNO to the $\chi^2$ is given by
$\chi^2_{SNO}=\chi^2_{gl,SNO}+\chi^2_{spec,SNO}$ where  
$\chi^2_{spec,SNO}$ is the spectrum contribution made up by the
day and night  values of the total (NC+CC+ES) SNO signal for the 
different values of the spectrum \cite{Aliani:2002na}. 
The new component corresponding to the individual global signals
is given by
\begin{eqnarray}
  \chi^2_{\rm gl,SNO}
&=&\sum_{i=ES,CC,NC} ({\alpha \vR^{\rm th} -\vR^{\rm exp}})^t 
\left (\sigma^{2}_{\rm stat} + \sigma^{2}_{\rm syst}\right )^{-1}
 ({\alpha \vR^{\rm th}-\vR^{\rm exp}}),
\label{chiall}
\end{eqnarray}
where the signal  $\vR$ vectors of dimension 3 
are made up by the
values of the total ES, CC, NC SNO signals.
The statistical contribution to the covariance 
matrix, $\sigma_{\rm stat}$ is non-diagonal since the different 
fluxes are derived from a fit to a single data sample \cite{newSNO,snohowto}.
 The part of the matrix 
related to the systematical errors 
includes contributions from  neutron capture efficiency and other 
geometrical inefficiencies  appearing in the statistical separation 
of ES, CC and NC events as presented in Ref.\cite{snohowto}.

\section{Results and Discussion}

To test a particular oscillation hypothesis against the parameters 
of the best fit and obtain allowed regions in parameter space we perform a 
 minimisation of the three dimensional function
 $\chi^2(\Delta m^2,\tan^2\theta,\alpha)$. 
For $\alpha=\alpha_{\rm min}$, 
 a given point in the oscillation parameter space is allowed if 
 the globally subtracted quantity fulfills the condition 
 $\Delta \chi^2=\chi^2 (\Delta m^2, \theta)-\chi_{\rm min}^2<\chi^2_n(CL)$.
Where $\chi^2_{n=3}(90\%,95\%,...)$ are the quantiles for
 three degrees of freedom.

The results are shown in Figs.\ref{f1} where we have generated acceptance 
contours in the $\Delta m^2$-$\tan^2\theta$ plane. 
In Fig.\ref{f2} we present the same results as in Fig.\ref{f1} but
using linear scales. The resolution in the neutrino parameter space
has become good enough for this to become useful.
With the actual experimental precision we can assert that the mixing
parameters are now known much better than as order of magnitude
only.
In  Table~(\ref{t2}) we present the best fit parameters or local minimum 
 obtained from the minimisation of the full $\chi^2$ function.

The main difference with previous analysis is a better resolution in 
parameter space. 
The previously 
two well separated solutions  LMAI,LMAII have now disappeared.
In particular the secondary region at larger mass differences
(LMAII) is now excluded at 95\% CL.

The introduction of the new solar data in general strongly 
diminishes the favored value for the 
mixing angle with respect to the KamLAND result alone \cite{Aliani:2002na}. 
The final value is  more near to those values favored by the solar data 
alone than to the KamLAND ones. 
As an important consequence, 
the combined analysis of solar and KamLAND data concludes that maximal mixing is 
not favored at $\sim 4-5\sigma$. 
This conclusion is not supported by the 
antineutrino, earth-controlled, conceptually simpler KamLAND results alone.
As we already pointed out in Ref.\cite{Aliani:2002na}, 
this effect could be simply due to the present low 
KamLAND statistics or, more worrying, to some statistical artifact derived from the complexity 
of the analysis  and of the heterogeneity of binned data involved. 

Additionally, we perform a second kind of analysis in order to obtain
concrete values for the individual oscillation parameters and
estimates for their uncertainties. 
We study the marginalised parameter constraints where the  $\chi^2$
quantity is converted into likelihood using the expression
${\cal L}/{\cal L}_0=e^{-(\chi^2-\chi_{min}^2)/2}$. 
This normalized marginal likelihood, obtained from the integration of ${\cal L}$ for each 
of the variables, is plotted in Figs.~(\ref{f3}) for each of the oscillation 
parameters $\Delta m^2$ and $\tan^2\theta$. 
Concrete values for the parameters are extracted by fitting  one- or two-sided 
Gaussian distributions to any of the peaks (fits not showed in the plots). In both cases, for  angle 
and the mass difference distributions the goodness of fit of the Gaussian fit to each individual peak 
is excellent (g.o.f $\sim 100\%$). The values for the 
parameters obtained in this way appear in Table~\ref{t2}. 
The errors obtained from this method are assigned to the $\chi^2$ minimisation values.
The central values are fully consistent and very similar to 
the values obtained from simple $\chi^2$ minimisation. 
In particular, the maximal mixing 
solution is again excluded at the $\sim 4-5\sigma$ level. 
A common feature to previous analysis presented by us \cite{Aliani:2002na} is that,
although both are mutually compatible, the slight difference of the value obtained 
for the mixing angle is well explained by the shape of the allowed regions in 
Fig~\ref{f1} (right): the right elongation of these shift the value of
the integral which defines 
the marginal distribution for $\tan^2\theta$. Additional variability can 
be easily 
introduced if would have used different prior information or mixing parameterizations.  

We will again use the technique of marginal distributions in the next
paragraphs to obtain an estimation of the individual elements of the
neutrino mass matrix and their errors.

\subsection{An estimation of the neutrino mass matrix}

The square of the neutrino mass matrix can be written in the flavour basis as 
$M^2=U M_D^2 U^\dagger $
 where $M_D$ is diagonal and $U$ is an unitary (purely 
active oscillations are assumed) mixing matrix. Subtracting one of the diagonal 
entries  we have
$$ M^2=m_1^2 I+M_0^2= m_1^2 I + U M_D^{\prime 2} U^\dagger, $$
where $I$ is the identity matrix. In this way we distinguish in the mass matrix a 
part, $M_0^2$, which affects and can be determined by  oscillation experiments
and another one,  $m_1^2 I$, which does not. Evidently, 
the off-diagonal elements of the mass matrix  are fully measurable 
 by oscillation experiments.  

Restricting ourselves for the sake of simplicity to two neutrino
oscillations, we have
\begin{eqnarray}
M^2=m_1^2 I+M_0^2&=& m_1^2 I + \Delta m^2
\pmatrix{\sin^2\theta & \sin\theta \cos\theta \cr
\sin\theta \cos\theta & \cos^2\theta }
\end{eqnarray}
with $\Delta m^2=m_2^2-m_1^2$. The individual elements of the matrix $M_0$ can simply be
estimated from the oscillation parameters obtained before. For example for 
$\tan^2\theta\sim 0.40$, $\Delta m^2\sim 7\times 10^{-5}\ eV^2$ we
would obtain $(M_0^2)_{22}\sim 5\times 10^{-5}\ eV^2$.

Our objective is however to  estimate how well the individual errors of the 
mass matrix can be extracted already at present by the existing experimental 
evidence.
For this purpose we have applied similar arguments as those used
before  to obtain marginal distributions and errors for individual
parameters from them. Using again as 
likelihood  function the quantity ${\cal L}/{\cal L}_0(\Delta
m^2,\tan^2\theta)=e^{-(\chi^2-\chi_{min}^2)/2}$ we obtained 
the individual probability distributions  for any of the elements of
the matrix $M_0$. 
Average values and $1\sigma$ errors are obtained from two-sided
Gaussian fits to these distributions.

From this procedure we obtain:  
\begin{eqnarray}
M_0^2&=& 
10^{-5}\ eV^2 
\pmatrix{ 2.06^{+0.29}_{-0.31} & 3.15^{+0.29}_{-0.35}\cr  3.15^{+0.29}_{-0.35} & 4.60^{+0.56}_{-0.44}}.
\end{eqnarray}

One can go further supposing a concrete value for $m_1^2$ from elsewhere. If
we take $m_1^2 >> \Delta m^2$ then we can directly write the
mass matrix
\begin{eqnarray}
M&=&m_1 I+\frac{1}{2 m_1} M_0^2.
\end{eqnarray}
Supposing for example $m_1=1\ eV$,
\begin{eqnarray}
M&=& eV 
\pmatrix{ 1.00+1.03^{+0.15}_{-0.15}\ 10^{-5}& 1.60^{+0.15}_{-0.17}\ 10^{-5}\cr  1.60^{+0.15}_{-0.17}\ 10^{-5}& 1.00+4.60^{+0.56}_{-0.44}\ 10^{-5}}.
\end{eqnarray}
this is  the final two neutrino mass matrix which can be obtained from present 
oscillation evidence coming from solar and reactor neutrinos.

\section{Summary and Conclusions}\label{sec:conclusions}

In this work we have presented an up-to-date analysis of all available
 Solar neutrino evidence including latest SNO results with NaCl
 enhanced efficiency in the most simple framework.
The increasingly accurate direct measurement via the $NC$ reaction on
 deuterium of 
${}^{8}{\rm B}$ neutrinos combined with the $CC$ results have largely 
confirmed the neutrino oscillation hypothesis.

In a model independent basis,
We obtain the following values for the ratios:
$\Phi_{CC}/\Phi_{NC}=0.305^{+0.030}_{-0.024}$,
$\Phi_{CC}/\Phi_{ES}=0.691^{+0.150}_{-0.096}.$
 The fraction a  of oscillating neutrinos
 into active and sterile ones are computed to be:
\begin{eqnarray}
\sin^2\alpha&=&0.940^{+0.065}_{-0.060}, \quad \cos^2\alpha \lsim
0.12\; (1\sigma),
\end{eqnarray}
where the fraction of oscillating sterile neutrinos 
 $\cos^2\alpha\equiv 1-\sin^2\alpha$. 
Although slightly increasing,
the central value is still clearly below one, the only-active
 oscillations case.
The hypothesis of transitions to {\em only} sterile 
  neutrinos is well rejected but, as a consequence of the reduced
 error bars, this data 
can be taken as a mildly positive hint in favour of a small sterile component:
the pure active case is now one sigma away from the central value.

We have  obtained the allowed area in parameter space 
and individual 
values for $\Delta m^2$ and $\tan^2\theta$ with error estimation
 from the analysis of marginal likelihoods. 
We have shown that it is already possible to 
determine at present active two neutrino oscillation parameters with
relatively good accuracy.
In the framework of two active neutrino oscillations we obtain 
$$
 \Delta m^2= 7.01\pm 0.08\times 10^{-5} \eV^2,\quad \tan^2\theta=
 0.42^{+0.12}_{-0.07}.
$$ 
The combined analysis of solar and KamLAND data concludes that maximal
mixing is not favored at $\sim 4-5\sigma$. This conclusion is not supported 
by the antineutrino, earth-controlled, conceptually simpler KamLAND
results alone.

We estimate the individual elements of the two neutrino mass matrix,
we show that individual elements of this matrix can be determined  
with an error $\sim 10 \% $ from present experimental evidence.

\vspace{0.3cm}
\subsection*{Acknowledgments}
It is a pleasure to thank R. Ferrari for many enlightening discussions and 
  for his encouraging and material support.
We  acknowledge the  financial  support of 
 the Italian MIUR, the  Spanish CYCIT  funding agencies 
 and the CERN Theoretical Division.
P.A. acknowledges funding from the Inter-University Attraction Pole (IUAP) 
"fundamental interactions".
The numerical calculations have been performed in the computer farm of 
 the Milano University theoretical group.

\newpage

\TABLE{
\centering
\scalebox{0.9}{
\begin{tabular}{|l|c|c|}
 \hline\vspace{0.1cm}
Experiment [Ref.] & $S_{Data}$ & $S_{Data}/S_{SSM}\ (\pm 1\sigma)$
\\[0.1cm]\hline 
New SNO data\protect\cite{newSNO}:  & & \\[0.2cm]
\ESSNO     & $ 2.21\pm 0.28\pm 0.10$ & $0.406\pm 0.091$ \\[0.1cm]
\CCSNO     & $ 1.59\pm 0.08\pm 0.07$ & $0.292\pm 0.056$ \\[0.1cm]
\NCSNO     & $ 5.21\pm 0.27\pm 0.38$ & $0.958\pm 0.193$ \\[0.2cm]
Other Solar data:  & & \\[0.2cm]
$D_2O$ \ESSNO\protect\cite{Ahmad:2002ka,Ahmad:2002jz} & $ 2.39\pm
0.23\pm 0.12$ & $ 0.439\pm0.092$ \\[0.1cm]
$D_2O$ \CCSNO\protect\cite{Ahmad:2002ka,Ahmad:2002jz} & $ 1.76\pm
0.05\pm 0.09$ & $ 0.324\pm0.054$ \\[0.1cm]
$D_2O$ \NCSNO\protect\cite{Ahmad:2002ka,Ahmad:2002jz} & $ 5.09\pm
0.43\pm 0.45$ & $ 0.936\pm0.208$ \\[0.1cm]
SK      \protect\cite{Fukuda:2002pe}  & $2.32\pm 0.03\pm 0.08 $ & $ 0.451\pm 0.011$   \\[0.1cm]
Cl          \protect\cite{cl1999}  &  $2.56\pm 0.16\pm 0.16   $ & $0.332\pm 0.056$ \\[0.1cm]
SAGE        \protect\cite{sage}  &  $67.2\pm 7.0 \pm 3.2  $ & $0.521\pm 0.067$ \\[0.1cm]
GNO-GALLEX  \protect\cite{gno2000,gallex}  &  $ 74.1\pm 6.7\pm 3.5 $ & $0.600\pm 0.067$  \\[0.1cm]
\hline
\end{tabular}
}
\caption{\small
Summary of data used in this work. 
The observed signal ($S_{Data}$) and  ratios $S_{Data}/S_{SSM}$
 with respect to the BPB2001 model are reported.
The  SK  and SNO rates are in \fluxunit units.
The   Cl , SAGE and GNO-GALLEX measurements are in SNU units. 
In this work we use the combined results of SAGE and GNO-GALLEX: 
 $S_{\rm Ga}/S_{\rm SSM}$ ($\rm Ga\equiv SAGE$+GALLEX+GNO)=$0.579\pm 0.050 $.
The SSM \boron total flux is taken from the (revised) 
BPB2001 model~\protect\cite{bpb2001}:
 $\Phi_{\nu}(\boron)=5.44 (1^{+0.20}_{-0.16}) \times 10^6$ cm$^{-2}$ s$^{-1}$.
In addition we have used from  reactor Kamland measurements the  signal ratio 
$R=0.611\pm 0.085\pm 0.041$  \cite{Eguchi:2002dm}and its signal spectrum 
\protect\cite{Aliani:2002na,Eguchi:2002dm}.
}\label{t1}
}

\TABLE{
      \begin{tabular}{lll}
 & $\Delta m^2 (\eV^2) $& $\tan^2\theta$
\\[0.15cm]
\hline
From minimization $\chi^2$:                     &  $7.01\times 10^{-5}$                 & $0.42$   \\[0.15cm]
From Marg. Fit, ( $\pm 1\sigma$): & $7.30^{+0.08}_{-0.08}\times 10^{-5} $ &  $0.46^{+0.12}_{0.07}$   \\[0.15cm]
\hline
      \end{tabular}
 \caption{\small
Mixing parameters: from $\chi^2$ minimization,  $\chi^2/ndf=0.94$
      (Fig. \protect\ref{f1} right) and from double-sided fit to the
      peak of marginal likelihood distributions
      (Figs.\protect\ref{f3}).}\label{t2}
}

\FIGURE{
\centering
\begin{tabular}{rl}
\psfig{file=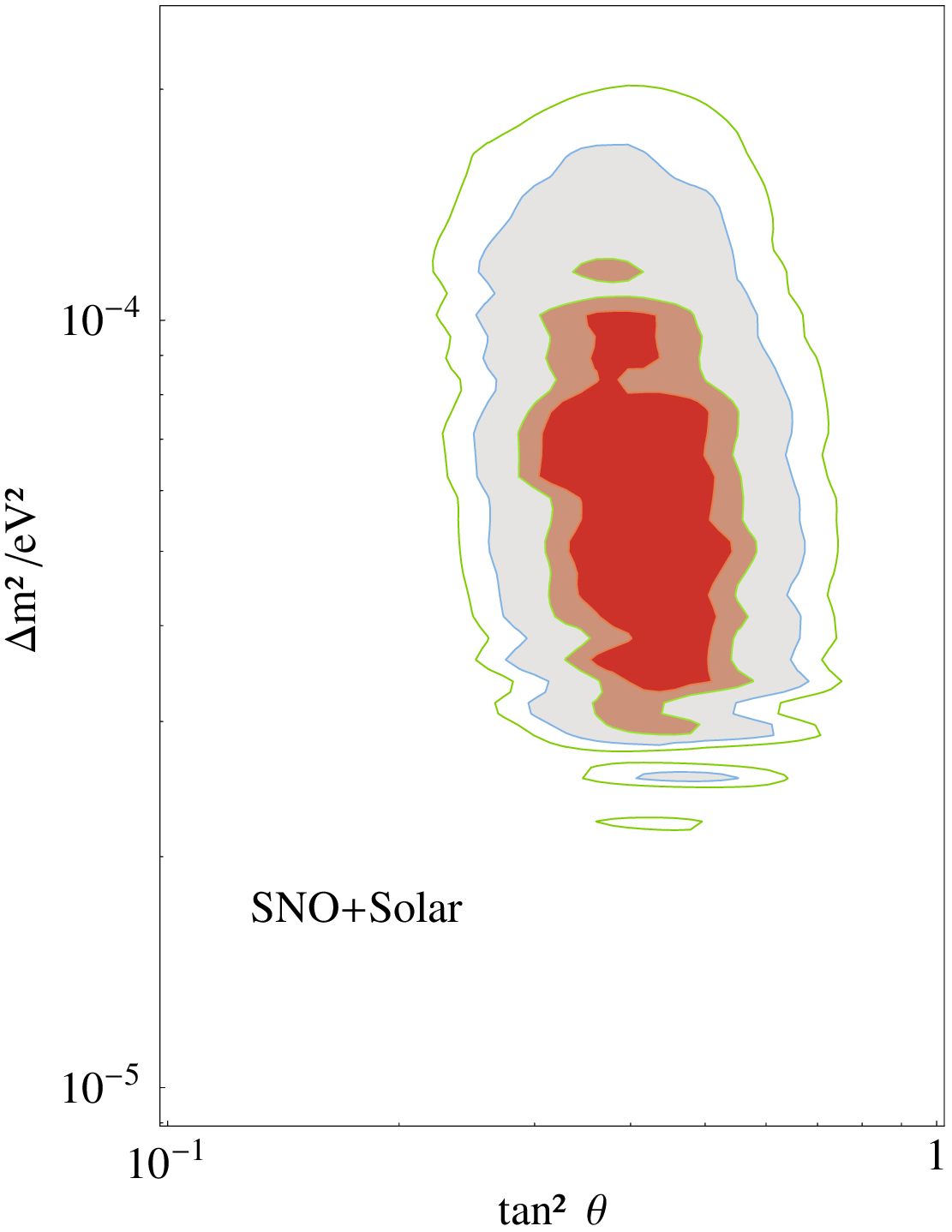,width=10cm,height=6.2cm} &
\hspace{-3cm}
\psfig{file=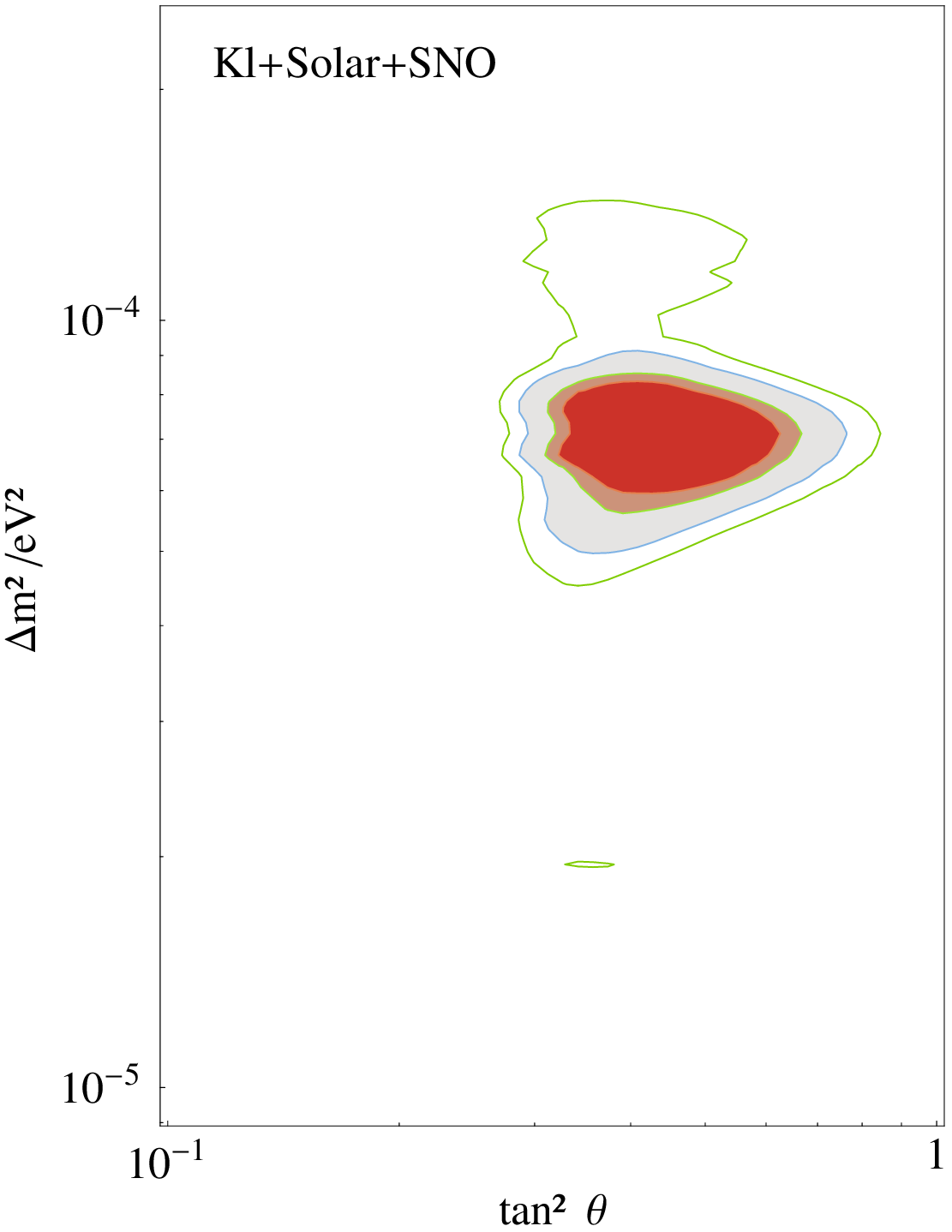,width=10cm,height=6.2cm}
\end{tabular}
\vspace{2cm}
\caption{\small
Allowed areas in the two neutrino parameter space.
The colored areas are the allowed regions at 
90, 95, 99 and 99.7\% CL relative to the absolute minimum.
(Left) Solar evidence (CL,GA,SK,SNO,SNO-salt).
(Right) Kamland spectrum plus solar  evidence.
}
\label{f1}
}

\FIGURE{
\psfig{file=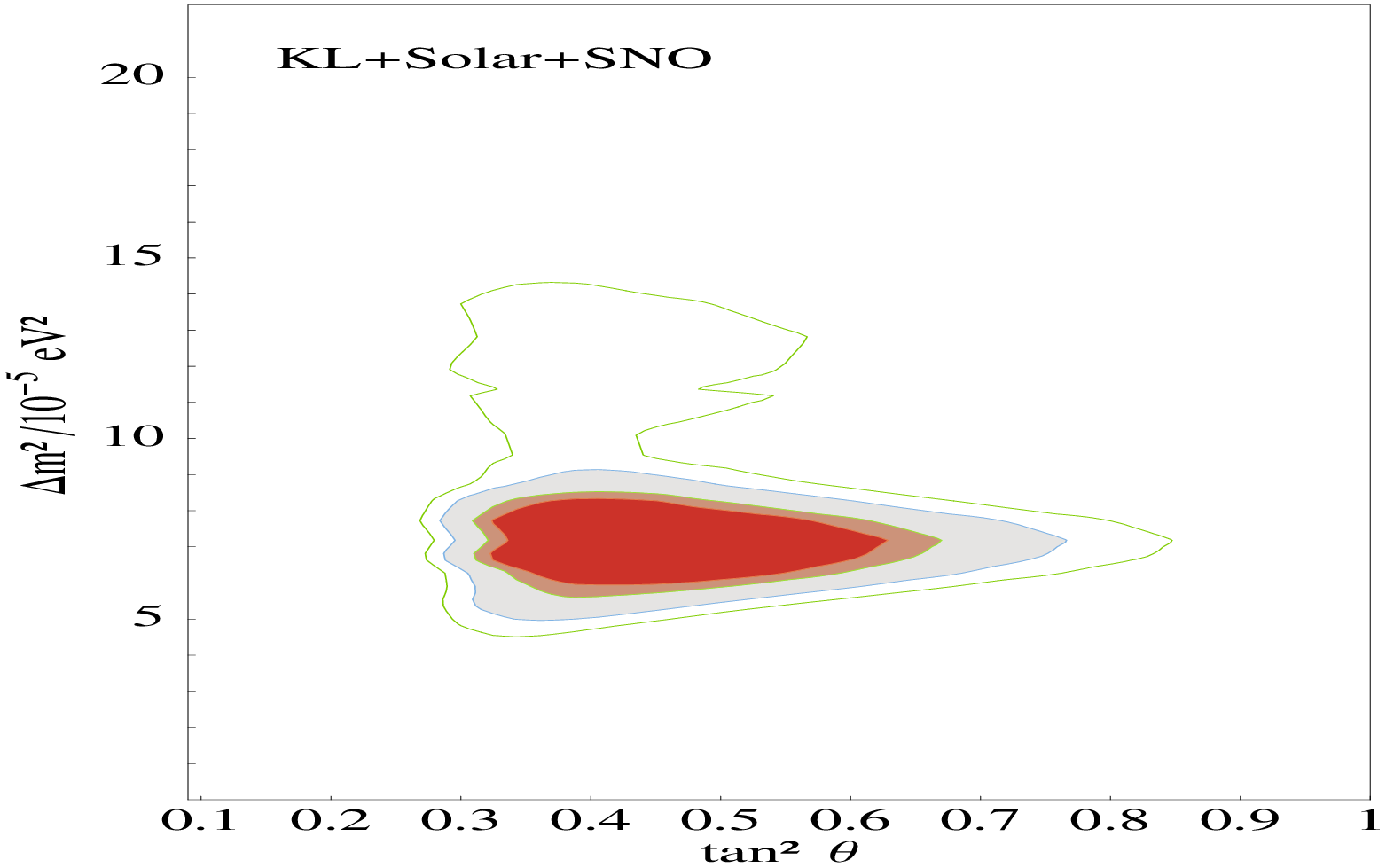,width=9cm,height=10cm}
\vspace{1cm}
\caption{\small
Kamland spectrum plus solar evidence as Fig.\protect\ref{f1} (right), in
linear scale which allows for a better comparison between the different
regions.} \label{f2}
}

\FIGURE{
\centering
\begin{tabular}{lr}
\hspace{-3.5cm}\psfig{file=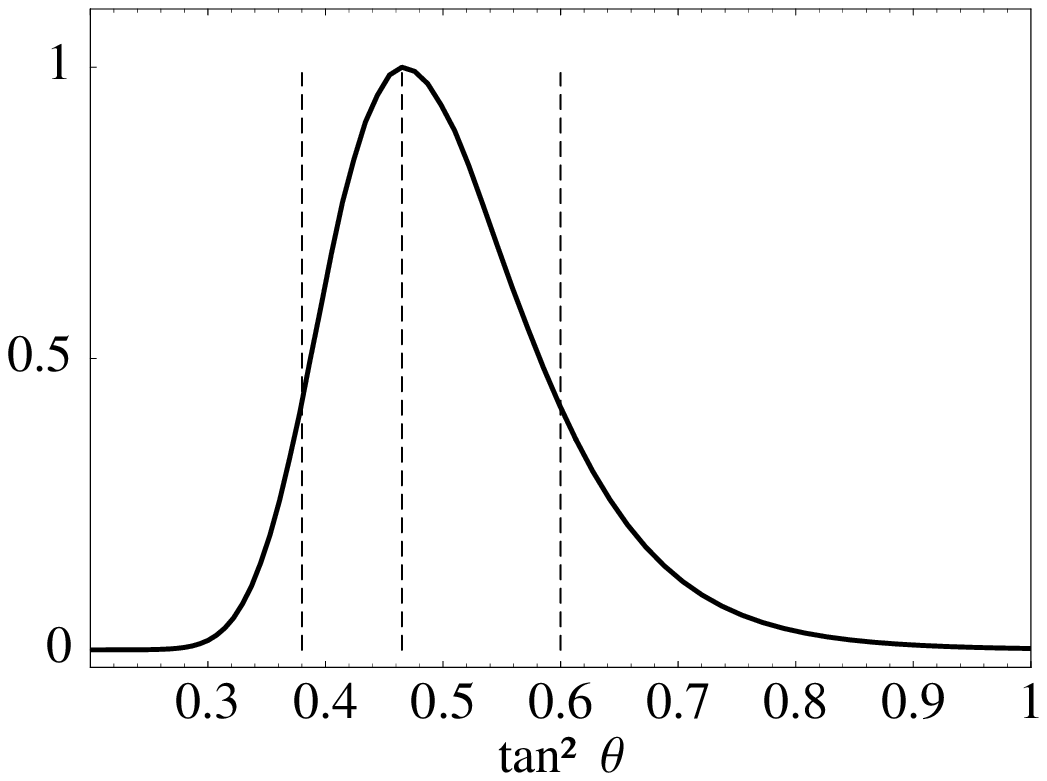,angle=0,width=7cm,height=7cm}&
\psfig{file=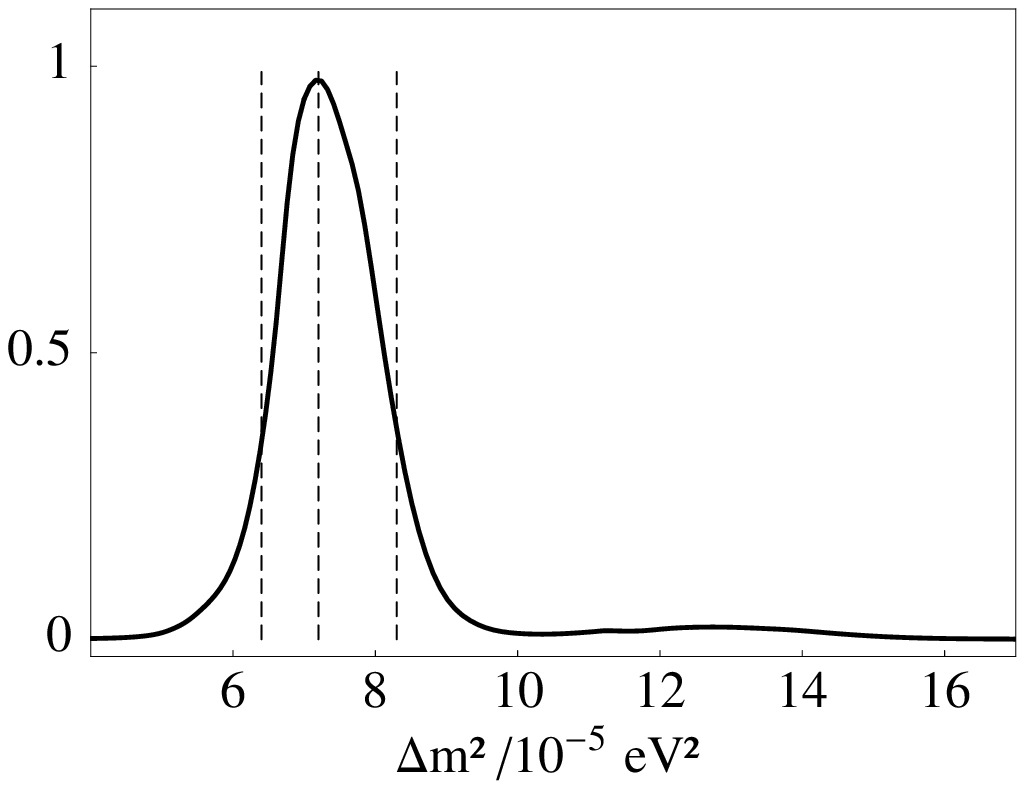,angle=0,width=7cm,height=7cm}
\end{tabular}
\vspace{-2.0cm}
\caption{\small  
Marginalized likelihood distributions for each of the
 oscillation parameters 
$\Delta m^2$ (right), $\tan^2 \theta$ (left)
corresponding to the solar plus Kamland  evidence
 (Fig.\protect\ref{f1}(Right)).
The curves are in arbitrary units with normalization to the 
 maximum height.
Values for the peak position are obtained by fitting 
two-sided Gaussian distributions (not showed in the plot).
 See Table 2
 for values of the position and widths of the peaks.} \label{f3}
}


\vspace{0.3cm}

{\small
\begin{thebibliography}{99}

\bibitem{newSNO}
S.~N. Ahmed {\it et al.},
ArXiv:nucl-ex/0309004, [see \texttt{http://www.sno.phy.queensu.ca}]

\bibitem{Fukuda:2002pe}
S.~Fukuda {\it et al.}  [Super-Kamiokande Collaboration],
Phys.\ Lett.\ B {\bf 539} (2002) 179.

\bibitem{Smy:2002fs}
M.~B.~Smy,
arXiv:hep-ex/0202020.

\bibitem{Ahmad:2002ka}
Q.~R.~Ahmad {\it et al.}  [SNO Collaboration],
Phys.\ Rev.\ Lett.\  {\bf 89}, 011302 (2002).

\bibitem{Ahmad:2002jz}
Q.~R.~Ahmad {\it et al.}  [SNO Collaboration],
Phys.\ Rev.\ Lett.\  {\bf 89}, 011301 (2002).

\bibitem{Eguchi:2002dm}
K.~Eguchi {\it et al.}  [KamLAND Collaboration],
{\em ``First results from KamLAND: Evidence for reactor anti-neutrino  disappearance,''}
Phys.\ Rev.\ Lett.\  {\bf 90}, 021802 (2003).

\bibitem{Migliozzi:2003pw}
P.~Migliozzi and F.~Terranova,
Phys.\ Lett.\ B {\bf 563} (2003) 73.

\bibitem{Bandyopadhyay:2003du}
A.~Bandyopadhyay, S.~Choubey and S.~Goswami,
Phys.\ Rev.\ D {\bf 67} (2003) 113011.

\bibitem{Bahcall:2003ce}
J.~N.~Bahcall and C.~Pena-Garay,
arXiv:hep-ph/0305159.

\bibitem{Choubey:2003qx}
S.~Choubey, S.~T.~Petcov and M.~Piai,
arXiv:hep-ph/0306017.

\bibitem{Aliani:2002na}
P.~Aliani, V.~Antonelli, M.~Picariello and E.~Torrente-Lujan,
Phys. Rev. D Vol. 68 (2003) 053000 [arXiv:hep-ph/0212212].

\bibitem{klothers}
G.~L.~Fogli, E.~Lisi, A.~Marrone, D.~Montanino, A.~Palazzo and A.~M.~Rotunno,
Phys.\ Rev.\ D {\bf 67} (2003) 073002;
V.~Barger and D.~Marfatia,
Phys.\ Lett.\ B {\bf 555} (2003) 144;
M.~Maltoni, T.~Schwetz and J.~W.~Valle,
Phys.\ Rev.\ D {\bf 67} (2003) 093003;
J.~N.~Bahcall, M.~C.~Gonzalez-Garcia and C.~Pena-Garay,
JHEP {\bf 0302} (2003) 009;
A.~Bandyopadhyay, S.~Choubey, R.~Gandhi, S.~Goswami and D.~P.~Roy,
Phys.\ Lett.\ B {\bf 559} (2003) 121;
W.~l.~Guo and Z.~z.~Xing,
Phys.\ Rev.\ D {\bf 67} (2003) 053002;
P.~C.~de Holanda and A.~Y.~Smirnov,
JCAP {\bf 0302} (2003) 001;
G.~L.~Fogli, E.~Lisi, A.~Marrone, D.~Montanino, A.~Palazzo and A.~M.~Rotunno,
arXiv:hep-ph/0308055;
T.~Schwetz,
arXiv:hep-ph/0308003;
H.~Nunokawa, W.~J.~Teves and R.~Zukanovich Funchal,
Phys.\ Lett.\ B {\bf 562} (2003) 28;
G.~Barenboim, L.~Borissov and J.~Lykken,
arXiv:hep-ph/0212116;
A.~Y.~Smirnov,
arXiv:hep-ph/0306075;
M.~C.~Gonzalez-Garcia and C.~Pena-Garay,
arXiv:hep-ph/0306001;
A.~Ianni,
J.\ Phys.\ G {\bf 29} (2003) 2107;
A.~B.~Balantekin and H.~Yuksel,
J.\ Phys.\ G {\bf 29} (2003) 665;
S.~Pakvasa and J.~W.~Valle,
arXiv:hep-ph/0301061.

\bibitem{Antonelli:2002zf}
V.~Antonelli, F.~Caravaglios, R.~Ferrari and M.~Picariello,
Phys.\ Lett.\ B {\bf 549} (2002) 325
[arXiv:hep-ph/0207347].
V.~Antonelli, F.~Caravaglios, R.~Ferrari and M.~Picariello,
arXiv:hep-ph/0305169.

\bibitem{ussnospec}
P.~Aliani, V.~Antonelli, M.~Picariello and
E.~Torrente-Lujan,
In preparation.

\bibitem{SNO}
J.~R.~Klein  [SNO Collaboration],
{\it  In *Venice 1999, Neutrino telescopes, vol. 1* 115-125}.
A.~B.~McDonald  [SNO Collaboration],
Nucl.\ Phys.\ Proc.\ Suppl.\  {\bf 77} (1999) 43.



\bibitem{Boger:2000bb}
J.~Boger {\it et al.}  [SNO Collaboration],
Nucl.\ Instrum.\ Meth.\ A {\bf 449} (2000) 172.

\bibitem{Barger:2001pf}
V.~Barger, D.~Marfatia and K.~Whisnant,
Phys.\ Lett.\ B {\bf 509} (2001) 19.

\bibitem{Bahcall:2001hv}
J.~N.~Bahcall, P.~I.~Krastev and A.~Y.~Smirnov,
JHEP {\bf 0105} (2001) 015.

\bibitem{newsnoothers}
G.~L.~Fogli, E.~Lisi, A.~Marrone and A.~Palazzo,
arXiv:hep-ph/0309100;
A.~B.~Balantekin and H.~Yuksel,
arXiv:hep-ph/0309079;
M.~Maltoni, T.~Schwetz, M.~A.~Tortola and J.~W.~Valle,
arXiv:hep-ph/0309130.
H.~Murayama and C.~Pena-Garay,
arXiv:hep-ph/0309114.

\bibitem{turck}
S.~Turck-Chieze,
Nucl.\ Phys.\ Proc.\ Suppl.\  {\bf 91} (2001) 73.
E.~G.~Adelberger {\it et al.},
Rev.\ Mod.\ Phys.\  {\bf 70} (1998) 1265.
A.~S.~Brun, S.~Turck-Chieze and P.~Morel,
arXiv:astro-ph/9806272.

\bibitem{bpb2001} 
J.~N.~Bahcall, M.~H.~Pinsonneault and S.~Basu,
Astrophys.\ J.\  {\bf 555}, 990 (2001).

\bibitem{bp95}
J.N. Bahcall and M.H. Pinsonneault, Rev. Mod. Phys. {\bf 67} (1995) 781.

\bibitem{sno2001}
Q.~R.~Ahmad {\it et al.}  [SNO Collaboration],
Phys.\ Rev.\ Lett.\  {\bf 87} (2001) 071301.

\bibitem{homestake}
R. Davis, Prog. Part. Nucl. Phys. 32 (1994) 13;
B.T. Cleveland et al., (HOMESTAKE Coll.) \npbps{38}{95}{47};
B.T. Cleveland et al., (HOMESTAKE Coll.)    Astrophys. J. 496 (1998) 505-526.

\bibitem{newgallium}
V. Gavrin, 4th International Workshop on Low Energy and Solar Neutrinos, 
Paris, May19-21,2003 ;                          
T. Kirsten, {\it Progress in GNO}, XXth Int. Conf. on Neutrino Physics and
Astrophysics, Munich, May 25-30, 2002; to be published in \
{ Nucl. Phys. }(Proc. Suppl.){\bf B }  

\bibitem{Aliani:2002ma}
P.~Aliani, V.~Antonelli, R.~Ferrari, M.~Picariello and E.~Torrente-Lujan,
Phys.\ Rev.\ D {\bf 67} (2003) 013006
[arXiv:hep-ph/0205053].

\bibitem{Bahcall:1996bw}
Y.~Fukuda {\it et al.}  [Super-Kamiokande Collaboration],
Phys.\ Rev.\ Lett.\  {\bf 82}, 1810;
J.~N.~Bahcall, P.~I.~Krastev and E.~Lisi,
Phys.\ Rev.\ C {\bf 55}, 494 (1997);
J.~N.~Bahcall and E.~Lisi,
Phys.\ Rev.\ D {\bf 54}, 5417 (1996).

\bibitem{Barger:2001zs}
V.~Barger, D.~Marfatia and K.~Whisnant,
arXiv:hep-ph/0106207 

\bibitem{torrente2001}
P.~Aliani, V.~Antonelli, M.~Picariello and E.~Torrente-Lujan,
Nucl.\ Phys.\ B {\bf 634} (2002) 393;
P.~Aliani, V.~Antonelli, M.~Picariello and E.~Torrente-Lujan,
Nucl.\ Phys.\ Proc.\ Suppl.\  {\bf 110}, 361 (2002);
E.~Torrente Lujan,
Phys.\ Rev.\ D {\bf 53}, 4030 (1996).
P.~Aliani, V.~Antonelli, R.~Ferrari, M.~Picariello and E.~Torrente-Lujan,
arXiv:hep-ph/0206308.
P.~Aliani, V.~Antonelli, R.~Ferrari, M.~Picariello and E.~Torrente-Lujan,
arXiv:hep-ph/0205061.

J.~N.~Bahcall, E.~Lisi, D.~E.~Alburger, L.~De Braeckeleer, S.~J.~Freedman and J.~Napolitano,
Phys.\ Rev.\ C {\bf 54}, 411 (1996).

\bibitem{Fogli:2001vr}
G.~L.~Fogli, E.~Lisi, D.~Montanino and A.~Palazzo,
Phys.\ Rev.\ D {\bf 64} (2001) 093007.

\bibitem{Krastev:2001tv}
P.~I.~Krastev and A.~Y.~Smirnov,
arXiv:hep-ph/0108177.

\bibitem{Bahcall:2001zu}
J.~N.~Bahcall, M.~C.~Gonzalez-Garcia and C.~Pena-Garay,
JHEP {\bf 0108}, 014 (2001);
J.~N.~Bahcall, M.~C.~Gonzalez-Garcia and C.~Pena-Garay,
[arXiv:hep-ph/0111150]

\bibitem{Bandyopadhyay:2001fb}
A.~Bandyopadhyay, S.~Choubey, S.~Goswami and K.~Kar,
arXiv:hep-ph/0110307.

\bibitem{Choubey:2001bi}
S.~Choubey, S.~Goswami and D.~P.~Roy,
arXiv:hep-ph/0109017.

\bibitem{Bandyopadhyay:2001aa}
A.~Bandyopadhyay, S.~Choubey, S.~Goswami and K.~Kar,
Phys.\ Lett.\ B {\bf 519} (2001) 83.

\bibitem{torrente}
E.~Torrente-Lujan,
Phys.\ Rev.\ D {\bf 59} (1999) 093006.
E.~Torrente-Lujan,
Phys.\ Rev.\ D {\bf 59} (1999) 073001.
E.~Torrente-Lujan,
Phys.\ Lett.\ B {\bf 441} (1998) 305.
V.~B.~Semikoz and E.~Torrente-Lujan,
Nucl.\ Phys.\ B {\bf 556} (1999) 353.
E.~Torrente-Lujan,
Phys.\ Lett.\ B {\bf 494} (2000) 255.
E.~Torrente-Lujan,
arXiv:hep-ph/9902339.
S.~Khalil and E.~Torrente-Lujan,
J.\ Egyptian Math.\ Soc.\  {\bf 9}, 91 (2001)
[arXiv:hep-ph/0012203].

\bibitem{Aliani:2003me}
P.~Aliani, V.~Antonelli, M.~Picariello and E.~Torrente-Lujan,
arXiv:cs.ce/0307053.

\bibitem{earthprofile}  
I.~Mocioiu and R.~Shrock,
Phys.\ Rev.\ D {\bf 62} (2000) 053017;
A. Dziewonski, in {\em The Encyclopedia of Solid Earth Geophysics},
edited by D.E James (Van Nostrand Reinhold, New York 1989).

\bibitem{sirlin1994} J. Bahcall, M. Kamionkowski, A. Sirlin, Phys. Rev. D51 (1995) 6146;
M. Passera, Phys. Rev. D. {64} (2001) 113002 .

\bibitem{nakamura2001}
S.~Nakamura, T.~Sato, V.~Gudkov and K.~Kubodera,
Phys.\ Rev.\ C {\bf 63} (2001) 034617.

\bibitem{snohowto}
Q.~R.~Ahmad {\it et al.}  [SNO Collaboration],
``HOWTO use the SNO solar Neutrino Spectral data'', http://www.sno.phy.queensu.ca/sno. 

\bibitem{Nakahata:1999pz}
M.~Nakahata {\it et al.}  [SK Coll.],
Nucl.\ Instrum.\ Meth.\ A {\bf 421}, 113 (1999).

\bibitem{skthesis}
H. Ishino, Ph. D. thesis, University of Tokio,1999.
M. Nakahata et al. (SK Coll.), Nucl. Instrum. Methods 46 (1998) 301.

\bibitem{sakurai} N. Sakurai, Ph.D. Thesis, Dec. 2000 {\em 
Constraints of the neutrino oscillation parameters from 1117 day 
observation of soloar neutrino day and night 
 spectra in Super-Kamiokande}.

\bibitem{Fukuda:2001nj}
S.~Fukuda {\it et al.}  [SK Coll.],
Phys.\ Rev.\ Lett.\  {\bf 86}, 5651 (2001).

\bibitem{Murayama:2000iq}
H.~Murayama and A.~Pierce,
Phys.\ Rev.\ D {\bf 65} (2002) 013012.

\bibitem{vogel}
 P.~Vogel and J.~F.~Beacom,                                      
 Phys.\ Rev.\ D {\bf 60}, 053003 (1999).  

\bibitem{kltorrente}
P.~Aliani, V.~Antonelli, M.~Picariello and E.~Torrente-Lujan,
JHEP {\bf 0302} (2003) 025.
E.~Torrente-Lujan,
JHEP {\bf 0304} (2003) 054
[arXiv:hep-ph/0302082].
B.~C.~Chauhan, J.~Pulido and E.~Torrente-Lujan,
Phys.\ Rev.\ D {\bf 68} (2003) 033015
[arXiv:hep-ph/0304297].

\bibitem{Aliani:2002ca}
P.~Aliani, V.~Antonelli, M.~Picariello and E.~Torrente-Lujan,
New J.\ Phys.\  {\bf 5} (2003) 2
[arXiv:hep-ph/0207348].
P.~Aliani, V.~Antonelli, R.~Ferrari, M.~Picariello and
E.~Torrente-Lujan,
reactor  neutrino physics,''
AIP Conf.\ Proc.\  {\bf 655} (2003) 103
[arXiv:hep-ph/0211062].

\bibitem{PDG2002} ``Review of Particle Properties'',
K. Hagiwara et al. (Particle Data Group), Phys. Rev. D 66 (2002) 010001

\bibitem{klstony} G.Horton-Smith,
{\em Neutrinos and Implications for Physics 
Beyond the Standard Model,} 
Stony Brook, Oct. 11-13, 2002.
http://insti.physics.sunysb.edu/itp/conf/neutrino.html

\bibitem{sage}
A.I. Abazov et al. (SAGE Coll.), \prl{67}{91}{3332}.
D.N. Abdurashitov et al. (SAGE Coll.), \prl{77}{96}{4708}.
J.N. Abdurashitov et al., (SAGE Coll.),  Phys. Rev. {\bf C60} (1999) 055801; 
astro-ph/9907131.
J.N. Abdurashitov et al., (SAGE Coll.), \prl{83}{99}{4686}.

%
%

\bibitem{gno2000} M. Altmann et al. (GNO Coll.)
 Phys. Lett. B490 (2000) 16-26.

\bibitem{gallex}
P. Anselmann et al., GALLEX Coll., \plb{285}{92}{376}.
W. Hampel et al., GALLEX Coll., \plb{388}{96}{384}.
T.A. Kirsten,  Prog. Part. Nucl. Phys. 40 (1998) 85-99.
W. Hampel et al., (GALLEX Coll.) \plb{447}{99}{127}.
M. Cribier, \npbps{70}{99}{284}.
W. Hampel et al., (GALLEX Coll.) \plb{436}{98}{158}.
W. Hampel et al., (GALLEX Coll.)     \plb{447}{99}{127}.

\bibitem{cl1999}
 K. Lande (For the Homestake Coll.) Nucl. Phys. B(Proc. Suppl.)77(1999)13-19.

\bibitem{kamlandmonaco}
J. Shirai, ``Start of Kamland'', talk given at {\it Neutrino 2002}, 
XXth International Conference on Neutrino Physics and Astrophysics,  
May 2002, Munich, 
\texttt{http://neutrino2002.ph.tum.de}.
See also: 
P.~Alivisatos {\it et al.},
STANFORD-HEP-98-03.
\end{thebibliography}
}
\end{document}